# Nanolaminated $Al_2O_3$/$HfO_2$ dielectrics for silicon carbide based devices

Raffaella Lo Nigro[1,a)], Emanuela Schilirò[1)], Patrick Fiorenza[1)] and Fabrizio Roccaforte[1)]

[1]Consiglio Nazionale delle Ricerche, Istituto per la Microelettronica e Microsistemi (IMM), Strada VIII n5, 90121 Catania, Italy

[a)] Electronic mail: raffaella.lonigro@imm.cnr.it

**Abstract**

Nanolaminated $Al_2O_3$/$HfO_2$ thin films as well as single $Al_2O_3$ and $HfO_2$ layers have been grown as gate dielectrics by Plasma Enhanced Atomic Layer Deposition (PEALD) technique on silicon carbide (4H-SiC) substrates. All the three dielectric films have been deposited at temperature as low as 250°C, with a total thickness of about 30 nm and in particular, the nanolaminated $Al_2O_3$/$HfO_2$ films have been fabricated by alternating nanometric $Al_2O_3$ and $HfO_2$ layers. The structural characteristics and dielectric properties of the nanolaminated $Al_2O_3$/$HfO_2$ films have been evaluated and compared to those of the parent $Al_2O_3$ and $HfO_2$ single layers. Moreover, the structural properties and their evolution upon annealing treatment at 800°C have been investigated as preliminary test for their possible implementation in the device fabrication flow-chart. On the basis of the collected data, the nanolaminated films demonstrated to possess promising dielectric behavior with respect to the simple oxide layers.

# I. INTRODUCTION

High critical electric field, high electron mobility and high thermal conductivity are the main characteristics of the silicon carbide, which is one the principally investigated wide band gap semiconductors for many electronics applications.[1] Among the several polymorphs of this material,[2] the (0001)4H-SiC substrates are the most used for their superior electronics properties and investigated in order to fabricate a new generation of power devices, whose performances are expected to go well beyond the silicon limits.[3-5] From an applicative point of view, (0001)4H-SiC substrate for power device fabrication offers the possibility to form the same native oxide as in silicon (i.e., $SiO_2$), which can be used as gate insulator in metal-oxide-semiconductor field-effect transistors (MOSFETs).[6] However, considering the permittivity values of both 4H-SiC substrate and $SiO_2$ insulator (9.7 and 3.9, respectively), the electric field in the gate oxide of a 4H-SiC MOSFET is typically a factor of 2.5 higher than in the 4H-SiC drift layer.[1] In this context, high permittivity (high-κ) insulators could allow to overcome this limitation, enabling a better redistribution of the electric field between insulator and SiC and guaranteeing safer operation conditions for the insulator in high voltage applications.

Aluminum oxide ($Al_2O_3$) thin films are the most studied high-κ dielectrics, due to the good insulating properties and to its thermal and chemical stability. In particular, the dielectric constant value (κ~9), the relatively large band gap (~ 9 eV) and the high critical electric field (10 MV/cm) make the $Al_2O_3$ material a promising high-κ oxide, suitable to replace the traditional $SiO_2$ dielectric in several microelectronics devices.[7-9] On the other hand, hafnium oxide ($HfO_2$) has been also extensively studied as a potential alternative to silicon oxide $SiO_2$ especially due to its quite high dielectric constant, (κ~ 20), while the principal $HfO_2$ drawback is related to its not so wide band gap (5.7 eV).[10]

Hence, it is clear that the selection of a dielectric for gate insulation in wide band gap semiconductors based devices, is not straightforward, but many issues need to the considered, such as the dielectric constant, bandgap and band alignment, breakdown field, mechanical and thermal stability.[11-15] As an example, to reduce the leakage current through the dielectric, the barrier height must be sufficiently high in combination with the high dielectric constant. In this context, the choice of the $Al_2O_3/HfO_2$ system is motivated by the possibility to combine the complimentary characteristics of the two materials, i. e. the $Al_2O_3$ excellent chemical and thermal stability, large band gap (around 9 eV) and the $HfO_2$ high dielectric constant. Therefore, their nanolaminated structure composed of the two high-k oxides is a promising solution for enhancing thermal stability and sustain a high dielectric constant value.[16] Several nanolaminated structures have attracted great attention for the possibility of tuning their mechanical or physical properties and have been applied in differently specific applications.[17] However, in microelectronics,[18,19] the $Al_2O_3$-$HfO_2$ nanolaminated films seem to be the most promising system and they have been mainly studied to be used for silicon based microelectronics[20-24] as well as in gallium nitride power electronics devices.[25-29] Poor literature is present on the study of nanolaminated systems on silicon carbide substrate. In particular, to our best knowledge only evaluation of an $Y_2O_3/Al_2O_3$ multilayer[30] and of a simple $Al_2O_3/HfO_2$ bilayer[31] have been reported.

In this paper, the growth via Plasma Enhanced Atomic Layer Deposition (PEALD) of nanolaminated $Al_2O_3$-$HfO_2$ multilayers has been implemented on (0001) 4H-SiC substrates at low deposition temperature of 250°C. The implementation of ALD on 4H-SiC substrates is related to the peculiarities of the deposition mechanism. In particular, because of the self-limiting principle and of the layer-by-layer growth mode, ALD guarantees the conformal and uniform film deposition on large area, with precise thickness control at relatively low deposition temperatures.[32] The described ALD features play a fundamental role in several new generation industrial fields.

The structural properties of the $Al_2O_3/HfO_2$ nanolaminated layers have been investigated and compared to those of the analogously grown $Al_2O_3$ and $HfO_2$ single layers. The study of the possible advantages of nanolaminated system with respect to the single oxides films has been completed by evaluation of their thermal stability upon post deposition high temperature annealings as well as of their dielectric properties.

## II. EXPERIMENTAL

Trimethylaluminum (TMA) and tetrakis- dimethylamino hafnium (TDMAH) have been used to deposit $Al_2O_3$ and $HfO_2$ films, respectively. These metal precursors were purchased by Air Liquid as bubblers. The TMA was used at room temperature, since it is a liquid precursor, while the hafnium precursor is a solid at room temperature, and it needed to be heated at 75°C.[33] A nitrogen flow of 40 sccm was used as carrier gas to deliver the TMA and TDMAH from the bubbler to the reactor chamber. Depositions were carried out on a 5 nm thermally grown $SiO_2$ layer formed on (0001) 4H-SiC substrates. Prior to $SiO_2$ thermal growth, the 4H-SiC wafers were treated with piranha solution ($H_2SO_4:H_2O_2$=1:3 for 10 min) and with diluted hydrofluoric acid ($H_2O:HF$=10:1 for 5 min) to remove and to clean the surface from carbon contaminations and eventual impurities. After these cleaning treatments, the 4H-SiC substrates were subjected to a controlled dry oxidation process at 1150°C in dry $O_2$-atmosphere to grow a 5 nm thick $SiO_2$ layer.[34]

*Film deposition parameters*

Films were deposited on PE ALD LL SENTECH Instruments GmbH reactor. The ALD system is equipped with a remote capacitively coupled plasma (CCP) source. It is excited by a 13.56 MHz RF generator via a matchbox. The power supply was 200 W and the oxygen-gas-flow-rate was 150 sccm.

Depositions were carried out at 250°C with a reactor chamber pressure of 20 Pa. For the $Al_2O_3$ deposition, the pulse times of metal precursor and oxygen pulse were 0.06s and 1s, respectively. For the $HfO_2$ deposition, the pulse times of metal precursor and oxygen pulse were 0.1s and 5s, respectively. After each precursor pulse, the deposition chamber was purged with 40 sccm $N_2$ to remove un-reacted precursors.

The $Al_2O_3$/$HfO_2$ nanolaminated films were deposited by repeating a unit cycle for precise thickness control. In particular, the $Al_2O_3$/$HfO_2$ nanolaminated film consisted of twenty total nanolayers. Each $Al_2O_3$ or $HfO_2$ nanolayer has been obtained by five unit cycles, with a unit cycle comprising a precursor, purge, oxygen reactant and final purge.

The deposited nanolaminated films were annealed by rapid thermal annealing process, which was carried out at 800°C in $N_2$ atmosphere for 1 minute in a Jetfirst 150 JIPELEC furnace.

*Structural and Morphological Characterizations*

Microstructures were investigated by transmission electron microscopy (TEM) using a FEG-TEM JEOL 2010F. Morphological characterization was performed by atomic force microscopy (AFM), which was carried out on a Digital Instrument D3100 operating in air.

*Device fabrications*

The test-patterns for the electrical characterization were fabricated using top Ni/Au electrodes. On the deposited films, Ni/Au bilayer, consisting of 80 nm of each metal, were deposited by sputtering and patterned by standard optical lithography and lift-off technique, to fabricate circular diodes.

The dielectric constants were determined by capacitance voltage (C/V) measurements, which were carried out using a Karl-Suss probe station equipped with an Agilent B1500A parameter analyzer.

## III. RESULTS AND DISCUSSION

Laminated $Al_2O_3/HfO_2$ thin films as well as single $Al_2O_3$ and $HfO_2$ layers have been deposited by plasma enhanced atomic layer deposition (PE-ALD) processes. Simple oxide $Al_2O_3$ and $HfO_2$ layers have been considered as references to compare the physical properties of their combination as laminated stack. Some of the deposition parameters, which have been used for the growth of the two different films, possess common values and in particular they consisted in deposition temperature of 250°C, 200 W plasma power, 40 sccm nitrogen gas carrier for the two metal precursors and total deposition pressure of 20 Pascal. The other parameters, which needed to be different for the growth of the two $Al_2O_3$ and $HfO_2$ layers have been reported on Table 1.

Table 1. Specific deposition parameters for the growth of $Al_2O_3$ and $HfO_2$ thin films

| Film | Metal Precursor Pulse/Purge | $O_2$ Plasma Pulse/Purge | Number of Cycles |
| --- | --- | --- | --- |
| $Al_2O_3$ | 0.06s/2s | 1s/2s | 250 |
| $HfO_2$ | 0.1/5s | 5s/2s | 100 |

Both growth processes have not been carried out on the bare (0001) 4H-SiC substrate but on a 5 nm thermally grown $SiO_2$ layer, which have been formed on its surface. This strategy has been proven to be the most useful in order to deposit quite dense and compact films, because of the presence of a greater number of active sites for nucleation with respect to the bare surface of the silicon carbide.[34] The growth rates of the $Al_2O_3$ and $HfO_2$ processes have been evaluated to be 1.1 Å/cycle and 1.6 Å/cycle, respectively. The indicated growth rate values have been calculated using the deposition parameters showed on table 1. In our previous papers,[16,34,35] the growth rates of the two different $Al_2O_3$ and $HfO_2$ layers grown on silicon substrate have been determined by evaluation of film thickness through ellipsometric measurements and corroboration of the thickness value by TEM imaging. An accurate evaluation of the growth rates was carried out by plotting the growth rate vs the number of cycles and linear trends were found in both cases. In this work, the growth rates have been determined by TEM analysis and they are in agreement with those found for $Al_2O_3$ and $HfO_2$ PE-ALD processes on silicon substrate. Cross-section TEM images (Fig. 1) have been recorded to also investigate the structural properties of the two $Al_2O_3$ and $HfO_2$ single layers. In particular, $Al_2O_3$ as-deposited film possesses a uniform thickness of about 30 nm, it is amorphous and adherent to $SiO_2$/4H-SiC substrate. On the other hand, the $HfO_2$ layer, deposited using 100 cycles as shown on table 1, demonstrated to have a lower thickness value of 14-16 nm and despite the amorphous nature of the initial 10-12 nm layer, some small grains are visible on the film surface. The formation of crystalline grains is an already known phenomenon reported literature,[35] generating a polycrystalline structure beyond a critical thickness of about 10 nm.

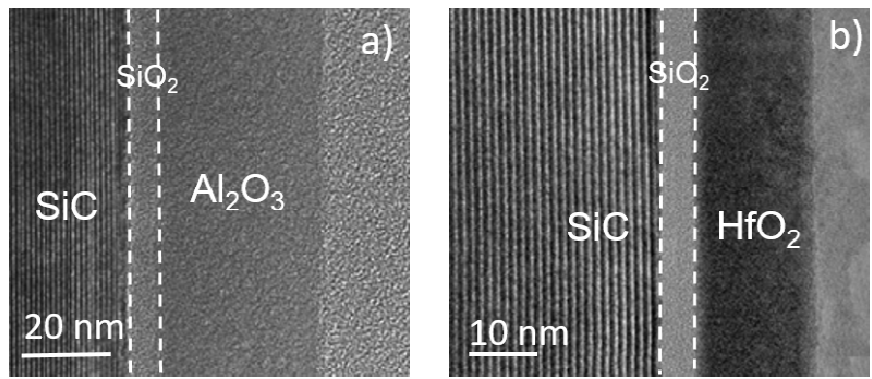

Fig. 1. TEM cross-section images of $Al_2O_3$ (a) and $HfO_2$ (b) thin films grown by PEALD on $SiO_2$/(0001) SiC substrates.

AFM characterization has been carried out to investigate the superficial morphology of the deposited film. In particular, 1μm×1μm areas have been imaged and the relative AFM bidimensional maps are shown Fig. 2 and compared to the substrate morphology (Fig. 2a). Film morphologies are characterized by smooth surfaces but also the presence of small circular grains is evident (Fig.s 2b and 2c). Their calculated root mean square (RMS) values, namely 0.56 nm and 0.61 nm, for $Al_2O_3$ and $HfO_2$ thin films, respectively, demonstrated that films are slightly rougher than the bare silicon carbide substrate (0.26 nm).

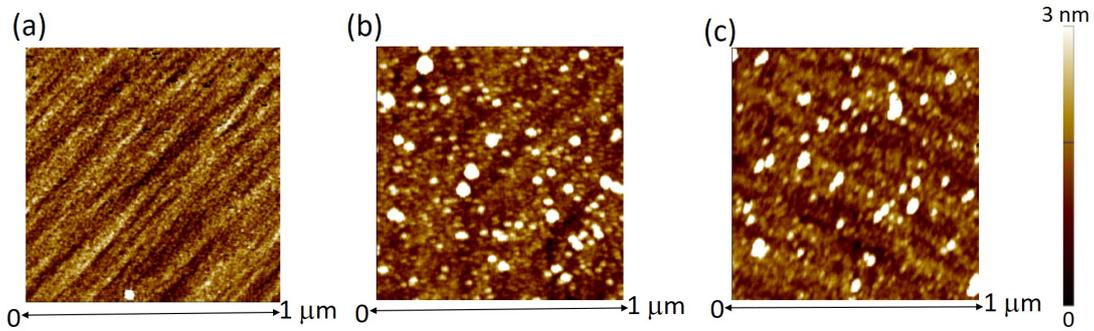

FIG. 2. AFM bidimensional maps of the SiO$_2$/4H-SiC substrate (a), Al$_2$O$_3$ (b) and HfO$_2$ (c) thin films grown by PEALD processes.

Finally, the dielectric properties of the two Al$_2$O$_3$ and HfO$_2$ films have been studied by the capacitance-voltage (C/V) measurements. The C/V curves (Fig. 3) of the two samples have been recorded on test-patterns which have been fabricated by several steps. C/V measurements have been performed at 1 MHz and the accumulation capacitances have been used to calculate the dielectric constant values. The dielectric constant value of the Al$_2$O$_3$ is 8.2, while in the case of HfO$_2$ films has been evaluated to be about 10.

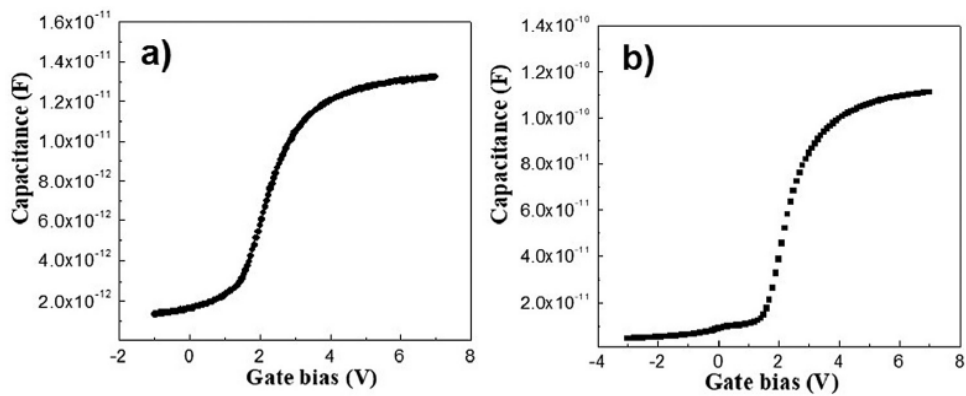

FIG. 3. C/V curves recorded at 1 MHz for Al$_2$O$_3$ (a) and HfO$_2$ (b) films deposited by PE-ALD processes.

The comparison of the characterization data between the two deposited films suggests a better thermal stability of the $Al_2O_3$ film with respect to that of $HfO_2$ layer, while this last possesses a higher dielectric constant value. The results are in agreement with the predicted behavior from the literature data[36] and their combination as laminated system has been considered in order to get the improved performances.

The proposed laminated system is schematically shown in Fig 4a and it consists of a 10 stacked $Al_2O_3$/$HfO_2$ bilayer, with each single oxide sub-layer having a thickness minor than 2 nm. The nanolaminated film has been fabricated using the same values of the main deposition parameters which have been described for the growth of the single layers, i. e. deposition temperature, process pressure and plasma value, while the cycles sequence has been appropriately modified. In particular, the two precursors have been alternatively pulsed into the reactor chamber and an extended purge time has been optimized in order to clean from the residual precursors both the substrate surface and the entire chamber, thus avoiding cross contamination between the different precursors.

The formation of the desired nanolaminated film has been demonstrated by TEM analysis, in fact, as shown in Fig 4c the formation of a laminated structure having sharp interfaces is clearly visible. The total film thickness is 36 nm and each sublayer is about 1.8 nm. The growth rate in the case of the nanolaminated film is about 1.2 Å/cycle and it is quite similar to the values of the processes for the growth of simple $Al_2O_3$ and $HfO_2$ oxides. In this context, it should be noted that in the TEM image, the thickness of the $HfO_2$ sub-layer seems to possess a slightly higher value than that of the $Al_2O_3$, however, this mis-interpretation could be due to the intrinsically darkest mass contrast of the $HfO_2$ with respect to the $Al_2O_3$ ones. An annealing treatment, at 800°C in $N_2$ atmosphere, has been performed on the laminated sample in order to check the thermal stability of the fabricated stack. The relative TEM image (Fig 4d) after the thermal treatment showed that it resulted in a small decrease (about 2 nm) of the total thickness, probably because of a densification process. In addition, the interfaces between the sub-layers became less sharp and diffusive phenomena, activated by high temperature, favored a light intermixing process between $Al_2O_3$ and $HfO_2$ nanolayers. Nevertheless, no structural evolution occurred upon the performed annealing process, in fact, no evident crystallization appeared as demonstrated by the electron diffraction pattern which did not show any diffraction spot except those from the substrate (not shown).

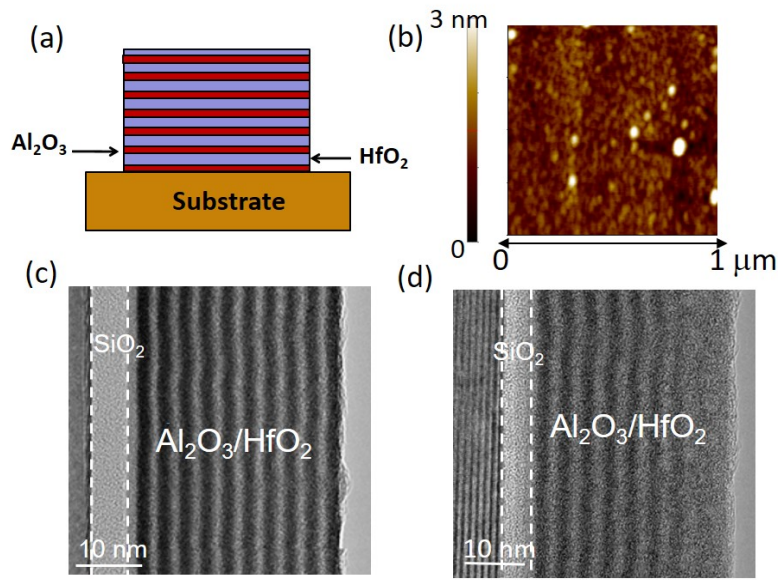

FIG. 4. Schematic of the $Al_2O_3/HfO_2$ laminated film (a) and its AFM bidimensional map (b). The cross section TEM image (c) shows the formation of about 1.8 nm alternating $Al_2O_3$ and $HfO_2$ thin layers in the as-dep film, whose interfaces became not so defined after annealing, as shown in the TEM image (d) recorded after the 800°C treatment.

This information on the amorphous nature of the annealed nanolaminated films represents an important result showing that the missed crystallization process, upon annealing at temperature well beyond the $HfO_2$ crystallization value (500°C), is a promising solution to improve the thermal stability of the nanolaminated dielectric.

In addition to the studies on the structural properties, the morphological characterization has been performed using AFM investigation. AFM mapping pointed out to a quite smooth surface and a morphology characterized by rounded grains, with root mean square of the height distribution value of 0.58 nm, quite similar to the values found in the single $Al_2O_3$ and $HfO_2$ single layers. Similar morphology as well as RMS value have been observed on the annealed samples.

A detailed electrical characterization of both as-deposited and annealed laminated samples has been performed by the capacitance voltage (C/V) curves, which have been reported in Fig 5. In particular, a more extensive study of the dielectrical properties has been carried out by evaluation of the shape and the position of the C-V characteristics which are directly related to the amount of charged defects in the insulator and at the insulator/semiconductor interface.

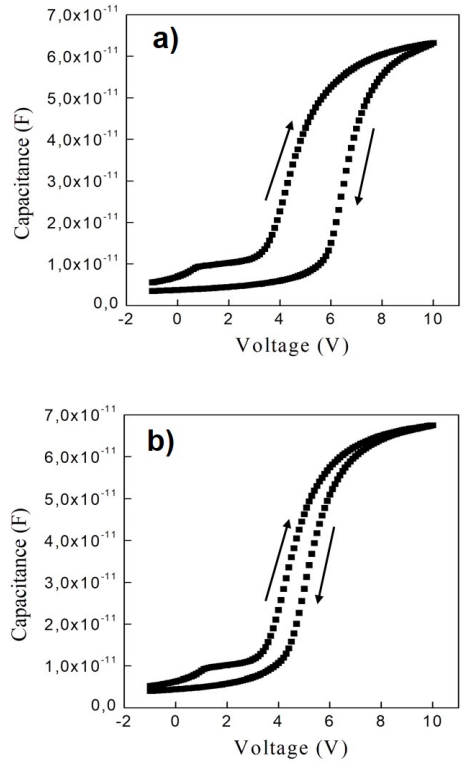

FIG. 5. C-V curves, recorded at 1 MHz, of $Al_2O_3/HfO_2$ nanolaminated films as-deposited by PE-ALD (a) and after annealing process at 800°C in $N_2$ atmosphere (b).

Hence, the C/V curves have been recorded at 1 MHz from depletion to accumulation and backwards, and have been used to experimentally determine not only the dielectric constant values but other dielectric parameters, such as the flat band voltage which is related to the number of charged defects and the hysteresis which is associated with the number of trapped (negative) charges.

The dielectric constant value of the as-deposited laminated film has been calculated to be 12.4.

The unbalanced charges arising from defects cause a C/V curve shift moving the flat band voltage from the ideal condition. In the case of Ni based contacts, the ideal flat band voltage value has been calculated to be $V_{FB\,Theor} = 1.28V$. The shift value between the theoretical ($V_{FBTheor}$) and experimental ($V_{FB}$) can be used to quantify the number of fixed charges ($N_f\,[cm^{-2}]$) in the dielectric film using the equation:

$$N_f = \frac{C_{ox}\Delta V_{FB}}{eA}$$

where $C_{ox}$ is the accumulation capacitance, $e$ is the elementary electron charge, $A$ is the area of the gate and $\Delta V_{FB}$ is the flat band shift.

In addition, by the flat band shift between the backward and forward C-V curves (hysteresis), the oxide trapping states ($N_{ot}$) can be quantified.

The estimated $N_f$ and $N_{ot}$ values are $N_f = 4\times10^{12}$ cm$^{-2}$ and $N_{ot} = 2.7\times10^{12}$ cm$^{-2}$, respectively.

After the annealing treatment at 800°C in $N_2$, the laminated stack showed an improvement of the dielectric properties. In fact, both the dielectric constant value improved to 13.4 and the $Q_{ot}$ value decreased to $1.15 \times 10^{12}$ cm$^{-2}$. The increasing of the dielectric constant value could be associated with the densification process already discussed and showed in the TEM image (Fig 4d), as well as it could be related to the probable formation of some single HfAlOx phase due to the light intermixing observed at the nanolayer interfaces.

All these results indicated a quite good insulating behavior of the laminated stack. Moreover, also the comparison of the complete structural and dielectric data of single $Al_2O_3$ or $HfO_2$ and their laminated combination, pointed out that the nanolaminated can be considered a promising system. In fact, it showed an amorphous structure before and after the annealing treatment and a better dielectric behavior in terms of dielectric constant and charge traps amounts. Further improvement of the dielectric properties could be achieved by chancing the $Al_2O_3$/$HfO_2$ ratio. In particular, maintaining constant the total film thickness and the number of alternating layer, while chancing the $Al_2O_3$ and $HfO_2$ nanolayer thicknesses, i.e. the $HfO_2$ thickness could be increased while the $Al_2O_3$ ones could be decreased. This strategy will be evaluated for our future experiments.

## IV. SUMMARY AND CONCLUSIONS

Huge efforts are nowadays devoted to the fabrication of multicomponent systems having high dielectric constants and good chemical stability. In particular, $Al_2O_3$ and $HfO_2$ oxides might be combined on the basis of their intrinsic physical properties. Three dielectric layers, namely $Al_2O_3$, $HfO_2$ and $Al_2O_3$/$HfO_2$ laminated, have been deposited on silicon carbide substrates by PEALD with the same thickness and their thermal stability and dielectric properties have been compared. The $Al_2O_3$ layer possesses good thermal stability but low dielectric constant. The $HfO_2$ layer possesses the higher dielectric constant value (about 10) but low thermal stability and as a consequence highest leakage current density and lowest reliability, while the combination of aluminum oxide and hafnium oxide in nanolaminated system improved the relatively low thermal stability of $HfO_2$ (~500°C) and increased the relatively low dielectric constant (~ 13) of $Al_2O_3$ thin films.

## ACKNOWLEDGMENTS

This work was funded by the European Community in the framework of the ECSEL Research and Innovation Actions (RIA) Calls 2016, Project WInSiC4AP (Wide band gap Innovative SiC for Advanced Power), Project Number: [737483]